\documentclass[amsmath,amssymb,11pt,tightenlines,superscriptaddress,reprint, notitlepage,aps,nofootinbib]{revtex4-1}
\pdfoutput=1 % if your are submitting a pdflatex (i.e. if you have
\usepackage{mathtools}
\usepackage[utf8]{inputenc}
\usepackage[english]{babel}
\usepackage{amsmath}
\usepackage{graphicx}
\usepackage{epsfig}
\usepackage{slashed}
\usepackage{amssymb}
\usepackage{color}
\usepackage{tabulary}
\usepackage{multirow}
\usepackage[normalem]{ulem}
\definecolor{MyDarkBlue}{rgb}{0.1, 0.1, 0.8} %defining the color 'MyDarkBlue'
\definecolor{SBlue}{rgb}{0.2, 0.4, 0.7} %defining the color 'MyDarkBlue'
\definecolor{MyLightBlue}{rgb}{0.22,0.51,0.9}
\definecolor{MyGreen}{rgb}{0.0, 0.5, 0.0}
\definecolor{BrickRed}{rgb}{0.8, 0.25, 0.33}
\RequirePackage{hyperref}
\hypersetup{colorlinks, citecolor=SBlue,linkcolor=MyDarkBlue, urlcolor=BrickRed}
\makeatletter

\newcommand\ovbb[0]{$0\nu\beta\beta$ }			  %0vbb
\newcommand\nme[0]{M}				  %NME
\newcommand\vecl[1]{{\bf{#1}}}         		  %vector letter
\newcommand\vecs[1]{\boldsymbol{#1}}     		  %vector symbol
\begin{document}

\title{\Large Flavoured Neutrinoless Double Beta Decay}

\author{Lukas Graf}
\email[E-mail: ]{lukas.graf@mpi-hd.mpg.de}
\affiliation{Max-Planck-Institut f{\"u}r Kernphysik, Saupfercheckweg 1, 69117 Heidelberg, Germany}

\author{Sudip Jana}
\email[E-mail: ]{sudip.jana@mpi-hd.mpg.de}
\affiliation{Max-Planck-Institut f{\"u}r Kernphysik, Saupfercheckweg 1, 69117 Heidelberg, Germany}

\author{Manfred Lindner} 
\email[E-mail: ]{manfred.lindner@mpi-hd.mpg.de}
\affiliation{Max-Planck-Institut f{\"u}r Kernphysik, Saupfercheckweg 1, 69117 Heidelberg, Germany}

\author{Werner Rodejohann}   
\email[E-mail: ]{werner.rodejohann@mpi-hd.mpg.de}
\affiliation{Max-Planck-Institut f{\"u}r Kernphysik, Saupfercheckweg 1, 69117 Heidelberg, Germany}

\author{Xun-Jie Xu}   
\email[E-mail: ]{xunjie.xu@mpi-hd.mpg.de}
\affiliation{Max-Planck-Institut f{\"u}r Kernphysik, Saupfercheckweg 1, 69117 Heidelberg, Germany}
\affiliation{Service de Physique Théorique, Université Libre de Bruxelles, Boulevard du Triomphe,
CP225, 1050 Brussels, Belgium}

\begin{abstract}
\noindent
We discuss a mechanism of neutrinoless double beta decay, where neutrinos of different flavours come into play. This is realized by effective flavour-violating scalar interactions.  As one consequence, we find that within the normal mass ordering the neutrino effective mass may no longer vanish due to contributions from other flavours. We evaluate the necessary nuclear matrix elements, consider the interference between the standard diagram and the new scalar one, and analyze a UV-complete model that realizes the scalar interaction. Tests of the complete model are possible at colliders and future neutrino experiments. 
Our scenario represents an alternative mechanism for neutrinoless double beta decay, where nevertheless lepton number violation resides only in Majorana mass terms of light neutrinos. 

\end{abstract}

\maketitle

%%%%%%%%%%%%%%%%%%%%%%%%%%%%%%%%%%%%%%%%%%%%%%%%%%

\section{\label{sec:intro}Introduction}
\noindent
Neutrinoless double beta (\ovbb$\!\!$) decay \cite{Furry:1939qr}, referring to the conversion of a nucleus 
\begin{align}
(A,Z) \rightarrow (A,Z+2) + 2e^-\,,
\end{align}
is of great interest to particle physics and cosmology  \cite{Rodejohann:2011mu,Deppisch:2012nb,Deppisch:2017ecm}. Its observation would imply the non-conservation of lepton number, a charge associated with an accidental global Abelian symmetry of the Standard Model (SM). It would also mean that a Majorana neutrino mass is generated at certain (possibly tiny) level,  hence,  neutrinos can be identified with their own antiparticles \cite{Schechter:1981bd,Duerr:2011zd}.  Given its importance,  an extensive worldwide experimental effort is being made to observe \ovbb decay \cite{DellOro:2016tmg,Dolinski:2019nrj}.  The current best experimental lower limits on \ovbb decay half-life are already above $10^{26}$ y \cite{KamLAND-Zen:2016pfg,Agostini:2020xta}.

\begin{figure}[t!]
    \centering
    \includegraphics[width=0.50 \textwidth]{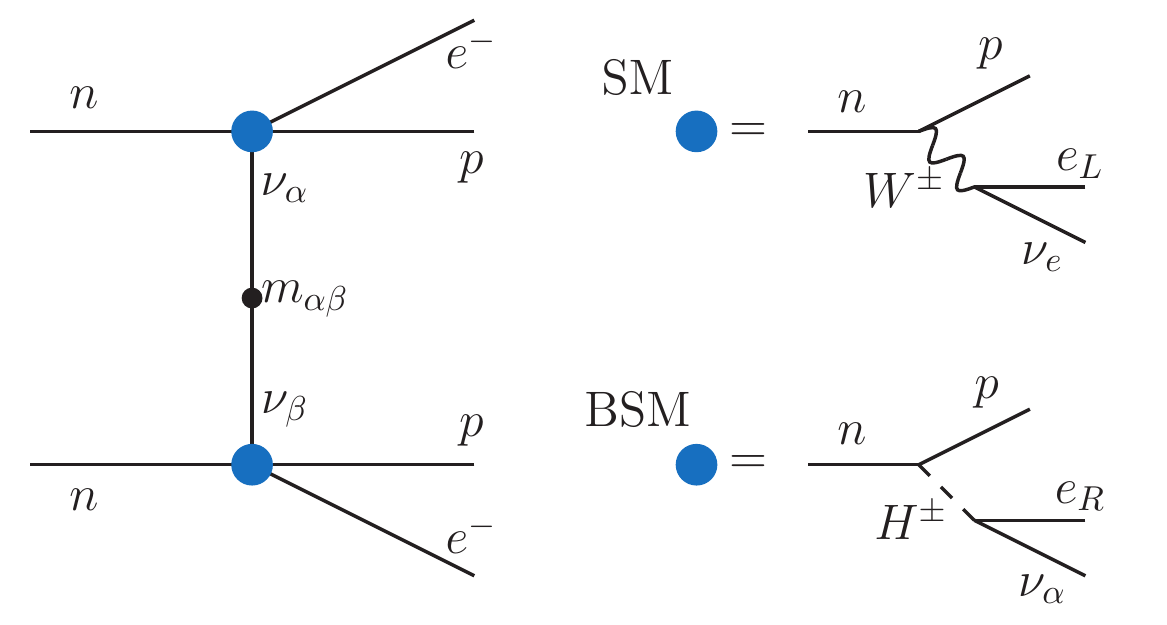}
    \caption{Flavoured double beta decay in and beyond the Standard Model. The standard diagram with mass-insertion $m_{ee}$ contains two SM-vertices generated by a  $W$-exchange. A new flavour-violating vertex generated by a charged scalar particle allows for a mass-insertion $m_{\alpha \beta}$.
    }
    \label{fig:feyn}
\end{figure}

In the standard interpretation of the decay, see Fig.\ \ref{fig:feyn}, internally at each $W$-electron vertex an electron neutrino appears, and the necessary spin flip renders the amplitude proportional to the \textsl{effective mass} 
% $U_{ei}^2 m_i$, where $m_i$ is the neutrino mass and $U_{ei}$ the PMNS matrix element relevant for the $W$-$e$-$\nu_e$ vertex. 
$U_{ei}^2 m_i \equiv m_{ee}$. Here $U_{\alpha i}$ are elements of the leptonic mixing matrix and $m_i$ the neutrino masses; this combination is furthermore nothing but the $ee$-element of the neutrino mass matrix. 
Since the $W$ boson mass is much larger than the nuclear scale, the standard diagram can be described by two  effective left-handed $u_L d_L e_L (\nu_e)_L$ vertices with a long-range Majorana neutrino exchange leading to a $m_{ee}$ mass-insertion, see Fig.\ \ref{fig:feyn}.  The parameters allow for a complete cancellation in case of a normal hierarchy, $m_{ee}=0$, which would lead to infinitely long lifetimes. This corresponds to the ``funnel" appearing in the usual plot of the effective mass in dependence on the lightest neutrino mass, see Fig.~\ref{fig:funnel}.

As commonly known, there exist several possibilities to generate the \ovbb decay with new physics \cite{Rodejohann:2011mu,Deppisch:2012nb}. We propose here the ``\textsl{flavoured}" version of the decay. That is, a mechanism with one or both of the left-handed $u_Ld_Le_L(\nu_e)_L$ vertices replaced by a beyond-the-Standard-Model scalar one, $u_Ld_Re_R(\nu_\alpha)_L$ (see Fig.\ \ref{fig:feyn}). In this case, a $m_{\alpha \beta} = U_{\alpha i} U_{\beta i} m_i $ mass-insertion can arise. This $u_Ld_Re_R(\nu_\alpha)_L$ flavour-violating, but lepton number conserving, vertex could originate from integrating out a charged scalar particle. 
The corresponding mass matrix elements $m_{\alpha \beta}$ have a different behavior as a function of the smallest neutrino mass and the mass ordering than $m_{ee}$ (see Fig.\ \ref{fig:mab}), hence the usual interpretation and phenomenology of the decay is altered.  Our scenario is distinctive from other alternative \ovbb mechanisms, as the new contribution to neutrinoless double beta decay depends on neutrino parameters. That is, lepton number violation occurs only in the Majorana mass terms of the light neutrinos.

Since the new contribution is suppressed by a mass-insertion and the constraint on the new scalar effective interaction, one may believe that its contribution is hopeless to observe. However, the nuclear matrix elements of the new contribution are enhanced, owing to a pion-exchange contribution, which somewhat compensates the double suppression. 

Indeed, we demonstrate in this paper that using this effective framework current limits from neutrino oscillation and other experiments allow for in principle observable half-lives of \ovbb decay mediated by the flavoured diagram, with numbers corresponding to  half-lives implied by meV-scale effective masses. Moreover, the different dependence on the neutrino parameters contained in $m_{\alpha \beta}$ means that the usual phenomenology of the decay is altered. For instance, within the normal mass ordering the lifetime may no longer be infinitely long due to parameter  cancellations, i.e.\ the ``funnel" gets sealed.

We then investigate options to generate the scalar $u_Ld_Re_R(\nu_\alpha)_L$ vertex with new physics in UV-complete models, focusing on a weak singlet charged scalar related to neutrino mass generation. This brings along additional experimental constraints, for instance from oscillation data or colliders.

\section{\label{sec:flav}Flavoured \ovbb Decay Rate}
\noindent 
Let us consider the following interactions
\begin{eqnarray}
{\cal L}_{{\rm int}} & = & {\cal L}_{V}+{\cal L}_{S}, \label{eq:m-0} \label{eq:lag} \\
{\cal L}_{V} & = & 2\sqrt{2}G_{F}V_{ud}\left[\overline{u}\gamma^{\mu}P_{L}d\right]\left[\overline{e}\gamma_{\mu}P_{L}\nu_{e}\right]+{\rm h.c.},\label{eq:lag_V}\\
{\cal L}_{S} & = & 2\sqrt{2}G_{F}\, \epsilon_{\alpha}V_{ud}\left[\overline{u}P_{L}d\right]\left[\overline{e}P_{L}\nu_{\alpha}\right]+{\rm h.c.}\label{eq:m}
\end{eqnarray}
Here ${\cal L}_{V}$ is the SM charged current (CC) interaction responsible for $0\nu\beta\beta$, and ${\cal L}_{S}$ contains a new 4-fermion non-standard interaction (NSI) of the scalar form, sometimes also referred to as a generalized neutrino interaction (GNI) \cite{Bischer:2019ttk}. 
Further, $G_F$ is the Fermi constant, $V$ is the CKM matrix, $P_L=(1-\gamma^5)/2$ is the left-handed projector, and $\epsilon_{\alpha}$ is a flavour-dependent coupling characterizing the strength of the new interaction. The projector in the quark sector in Eq.~(\ref{eq:m}) could be also right-handed,  but it would not affect the result significantly. 
We stress here that the operator in Eq.\ (\ref{eq:m}) conserves lepton  number. Therefore, lepton number violation is only triggered by the Majorana mass terms of the light neutrinos. This implies that our new contribution to neutrinoless double beta decay will depend on neutrino parameters, which distinguishes it from most other alternative mechanisms.

Given the above Lagrangian (\ref{eq:lag}), \ovbb decay is in principle generated by the standard diagram with two left-handed vector vertices (depending on $m_{ee}$), a diagram with one left-handed vector vertex replaced by the scalar one (depending on $m_{e\alpha}$), and a diagram with 
both vertices being of a scalar nature (depending on $m_{\alpha\beta}$). It turns out, however, that the mechanism combining the SM and BSM vertices vanishes at the leading order in our approximation; hence, we neglect it and focus only on the two other contributions. These are related to different nuclear matrix elements (NMEs), which we denote within this work as vector ($M_V$) and scalar ($M_S$). The two diagrams will interfere with each other, with the size of the interference term depending on the chirality of the electrons. The energies of the emitted electrons are larger than their mass by a factor of a few, which sets the order of magnitude of the interference term on the amplitude level.

Combining both the standard and the new physics contributions, the decay rate is given by (for details of the derivation, see Appendix \ref{app:a}) 

\begin{equation}  
    \Gamma_{0\nu} =
    \frac{G_{0\nu}}{m_{e}^{2}} \left|\tilde m_{ee}   M_V \right|^2 ,
    \label{eq:m-13}
\end{equation}
where $G_{0\nu}$ is the phase space factor, 
$M_V$ is the nuclear matrix element  of the standard $0\nu\beta\beta$ process, $m_e$ is the electron mass, and $\tilde m_{ee}$ is defined to include both the standard ($m_{ee}$) and flavoured ($m_{\alpha\beta}$) contributions as
\begin{align}
    \tilde{m}_{ee} &\equiv \left|m_{ee}\right|^{2}+\left|S\right|^{2} + 2\thinspace{\rm Re}\left[RSm_{ee}^{*}\right].\label{eq:mee_tilde}
\end{align}
\begin{figure}[t!]
    \centering
    
    \includegraphics[width=0.4\textwidth]{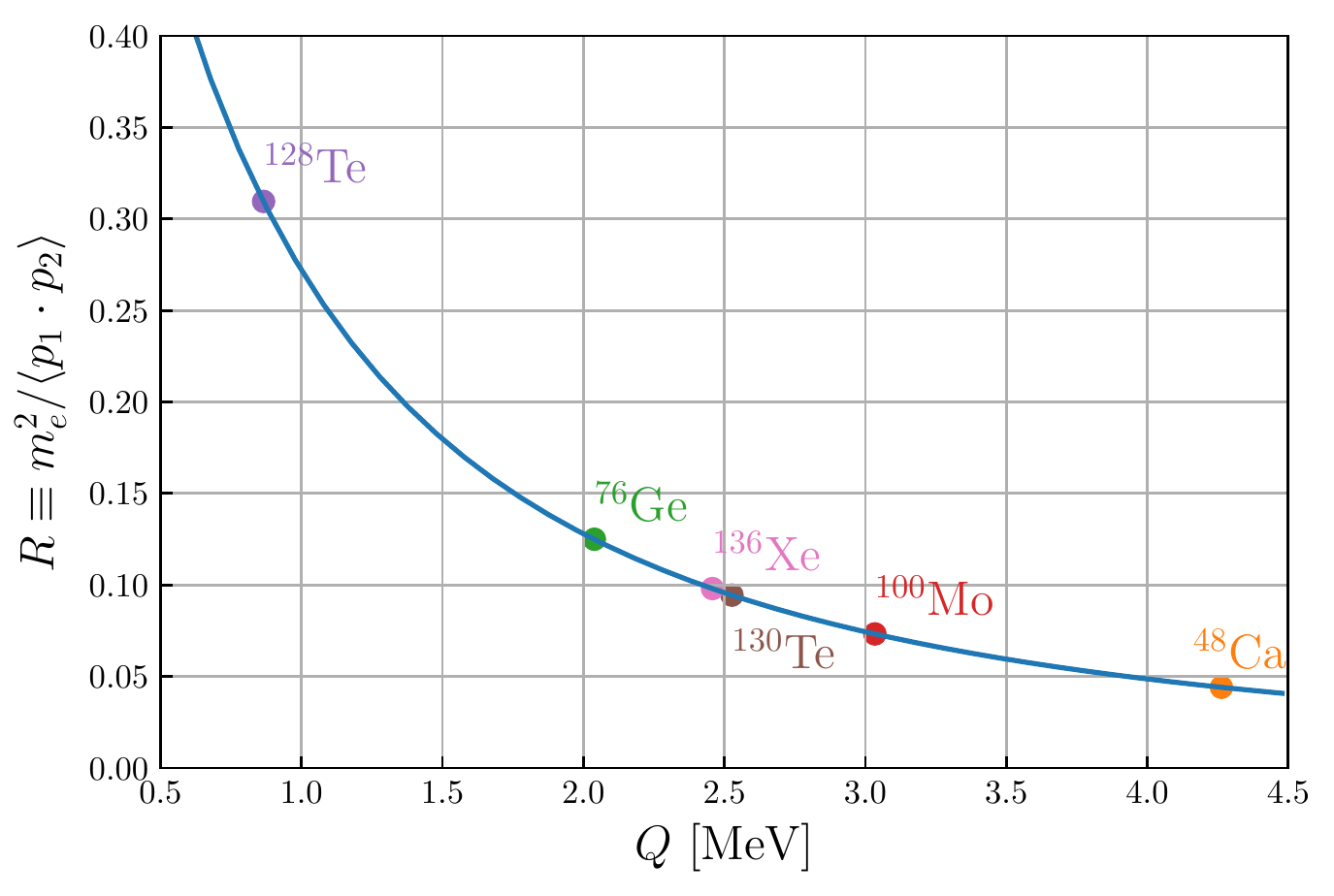}
    
    \caption{The interference factor $R$ is defined in Eq.~(\ref{eq:RmRE}) as a function of the $Q$ value.
    }
    \label{fig:Q}
    \end{figure}
Here we have defined combined quantity of particle and nuclear physics parameters, 
\begin{equation}
S\equiv m_{\alpha\beta}\thinspace\epsilon_{\alpha}\epsilon_{\beta}\frac{M_{S}}{M_{V}},
\label{eq:SX}
\end{equation}
and the interference factor of the diagrams
\begin{equation}
R\equiv\frac{\int_{p}m_{e}^{2}}{\int_{p}E_{1}E_{2}}. \label{eq:RmRE}
\end{equation}
The notation $\int_{p}$ stands for a phase space integral, 
\begin{equation}
\int_{p}F\equiv\int F\delta(\Delta M-E_{1}-E_{2})\frac{|\vecs{p_{1}}|^{2}|\vecs{p_{2}}|^{2} d |\vecs{p_{1}}| d|\vecs{p_{2}}|}{(2\pi)^{4}E_{1}E_{2}}, \label{eq:int_p}
\end{equation}
where we denote the two out-going electron momenta by $\vecs{p_1}$ and $\vecs{p_2}$,  their energies by $E_{1,2}=\sqrt{|\vecs{p_{1,2}}|^2+m_e^2}$ and the mass difference between the initial and final nuclei by $\Delta M$, which is connected to the $Q$ value as $\Delta M= Q+2m_{e}$. 
In Fig.\ \ref{fig:Q} we show $R$ for a variety of relevant isotopes.

Considering the NMEs, as mentioned in the introduction, the scalar ones enjoy an enhancement due to the pionic resonance manifesting through a large pseudoscalar nucleon form factor, which can be understood given the structure of these NMEs discussed in detail in  Appendix~\ref{app:b}. The numerical values of the total NMEs obtained for ${}^{76}$Ge and ${}^{136}$Xe  
using the Interacting Boson Model (IBM-2)~\cite{Barea:2009zza, Barea:2013bz, Barea:2015kwa} are summarized in Table \ref{tab:totnmes}. \\

\renewcommand{\arraystretch}{1.3}
\setlength{\tabcolsep}{12pt}
\begin{table}[t]
\centering
\begin{tabular}{lrrr}
\toprule
NME & $M_V$ &$M_S$ \\
\colrule
${}^{76}$Ge & -6.46 & 131  \\
${}^{136}$Xe & -3.57 & 70.5 \\
\botrule
\end{tabular}
\caption{\label{tab:totnmes}
Numerical values of the nuclear matrix elements obtained using the IBM-2 nuclear structure framework~\cite{Barea:2009zza, Barea:2013bz, Barea:2015kwa} for the standard and the flavoured \ovbb decay mechanisms. Note the large numbers for the NME $M_S$ corresponding to the mechanism involving two copies of the  scalar interaction, which are a consequence of the enhancement by the pionic resonance involved.
}
\end{table}

For an explicit example we need the experimental limits on the scalar coupling $\epsilon_{\alpha}$. 
At first, note that in neutrino oscillation experiments, this scalar interaction does not influence neutrino propagation but can cause flavour transition at zero distance.  In addition, CKM unitarity and leptonic universality also provide strong bounds. According to Refs.~\cite{Biggio:2009nt,Farzan:2017xzy}, the strongest bounds for $e$ and $\tau$ flavours  come from CKM unitarity, and for the $\mu$ flavour from neutrino oscillation experiments. Also note that in Refs.~\cite{Biggio:2009nt,Farzan:2017xzy}, only bounds on vector interactions (i.e., $ \epsilon_{\alpha}^V \left[\overline{u}\gamma^{\mu}P_{L}d\right]\left[\overline{e}\gamma_{\mu}P_{L}\nu_{\alpha}\right]$) were provided.
Properly translated in our $\epsilon_\alpha$ (overall rates  differ  by a factor $2$ due to the different Lorentz structure), the bounds 
are $\epsilon_{\mu} < 0.052$  and $\epsilon_{\tau} < 0.082$ at 90\% CL. 
There is also a more updated analysis on the bound from beta decay~\cite{Gonzalez-Alonso:2018omy} (the bound from CKM unitarity in Ref.~\cite{Biggio:2009nt} is based on beta decay combined with meson decay data), which reads $\epsilon < 0.063$ at 90\% CL for right-handed neutrinos with scalar interactions. Since for the case of muon and tau neutrinos interference with the SM processes is absent, this bound can be directly interpreted as a bound on $\epsilon_{\mu}$ and $\epsilon_{\tau}$. Combining  the bounds in the literature, we conclude that the current bounds on $\epsilon_{\mu}$ and $\epsilon_{\tau}$ are
\begin{equation}
    \epsilon_{\mu}<0.052\thinspace,\  \epsilon_{\tau}<0.063\ (90\%\ {\rm CL})\thinspace. \label{eq:m-1}
\end{equation}
\begin{figure}[t]
\centering

\includegraphics[width=0.5\textwidth]{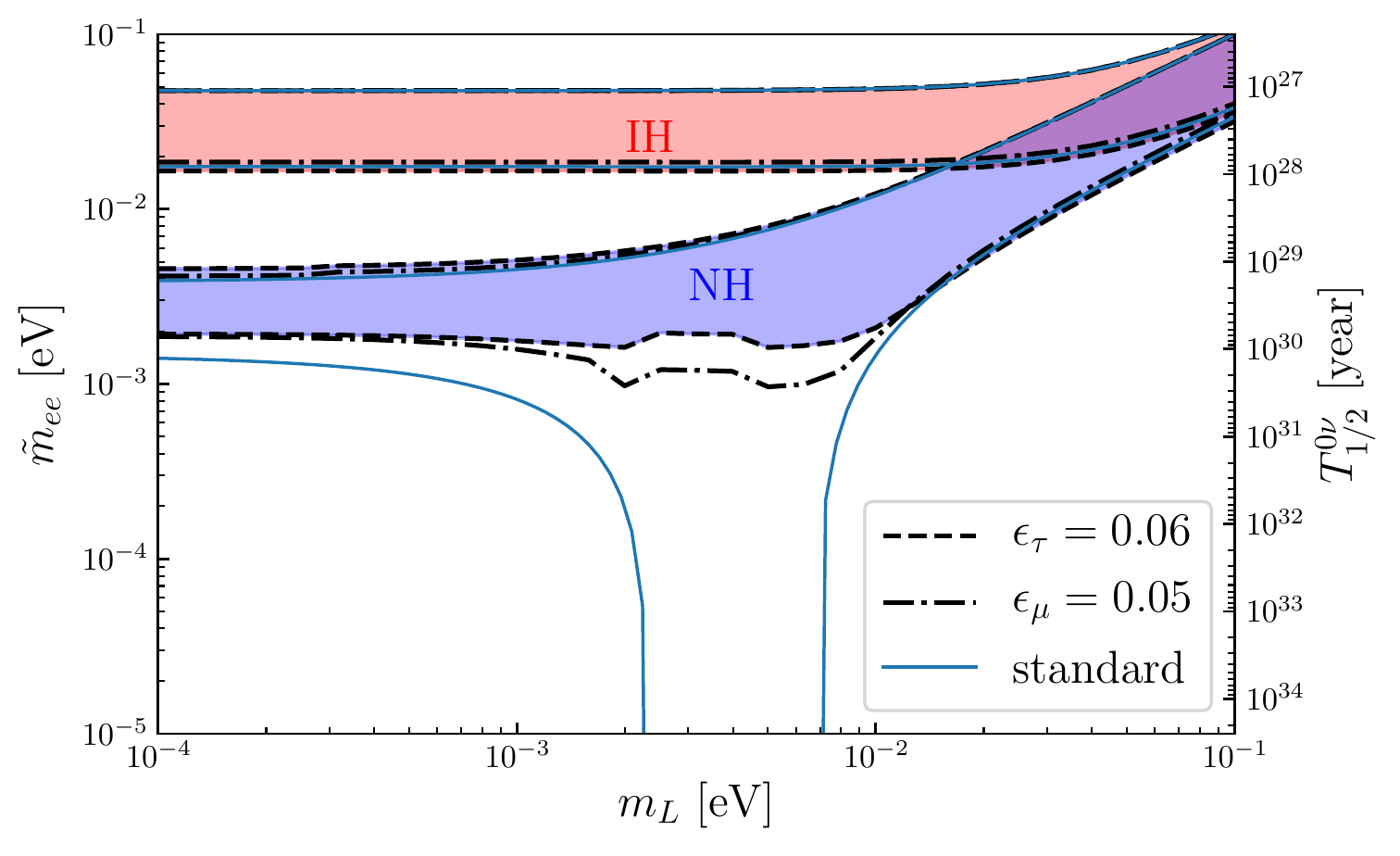} 
\caption{The effect of scalar interactions on $0\nu\beta\beta$. 
Here, $m_{L}$ is the lightest neutrino mass and $\tilde{m}_{ee}$ is defined in Eq.~(\ref{eq:mee_tilde}). 
The solid curves represent the standard $0\nu\beta\beta$ without
new interactions; the other curves include the new physics contributions
($\epsilon_{\tau}=0.06$ (dashed) and $\epsilon_{\mu}=0.05$ (dash-dotted)). We took here the isotope $^{76}{\rm Ge}$, the interference factor of which is $R=0.12$.
\label{fig:funnel}
} 
\end{figure}

As for $\epsilon_{e}$, due to interference in beta decay, the bound is much more stringent and it is also more complicated to convert the beta decay bounds on the vector $\epsilon_{e}^V$ to the scalar one because in this case the chirality flipping of $e_L$/$e_R$ is involved. Here we simply ignore the possibility of non-zero $\epsilon_{e}$, which would simply be a minor rescaling of the $0\nu\beta\beta$ decay half-life. 

For non-zero $\epsilon_{\mu}$ and $\epsilon_{\tau}$, \ovbb decay depends on $m_{\mu\mu}$ and $m_{\tau\tau}$, respectively. In case of $\epsilon_{\mu}$ and $\epsilon_{\tau}$ both being non-zero, $m_{\mu\tau}$ contributes in addition to $\tilde{m}_{ee}$. 
Of course, the standard diagram depending on $m_{ee}$ is also present.  
Using the values in Eq.~\eqref{eq:m-1}, we show in Fig.~\ref{fig:funnel} the (modified) effective mass as a function of the smallest neutrino mass $m_L$. We stress again the funnel for $m_L\in$ [0.002, 0.007] eV, where $m_{ee}$ vanishes due to cancellations. However, for our example,  $\tilde{m}_{ee}$ can not vanish anymore. This is due to the different behavior of $m_{\alpha\beta}\neq m_{ee}$ as a function of the smallest neutrino mass, cf.\ Fig.\ \ref{fig:mab}. The usual funnel is now sealed. 
Note that we also include the corresponding half-life of the decay in the plot. 
The numbers are large, as they correspond to meV-scale effective masses. The NMEs that we have calculated (for details see Appendix \ref{app:a}) use the IBM-2 approach. Those matrix elements are known to lead to larger half-lifes than some of the other approaches such as those based on the Energy Density Functional (see \cite{Engel:2016xgb} for a comparison), so that one could easily obtain smaller values. There are however also approaches which would lead to larger half-lifes, such as the Shell Model. This is of course the usual dilemma in calculations of NMEs. Note further that we choose a quenched value of $g_A=1$ for the axial coupling. 

We stress here the importance of the scalar nature of the new interaction. Its effect is the reduction of the interference term, which means that a complete cancellation of the standard and new physics diagram is not possible.

\section{UV completion}
\noindent
Let us now analyze an explicit realization of the effective scalar interaction. As a prototypical example, we consider the Zee model~\cite{Zee:1980ai} -- one of the most popular neutrino mass models, where neutrino masses and mixings are generated via quantum corrections at one-loop level, while NSIs are induced at tree level via exchange of charged scalar particles. The particle content of the Zee model contains an  $SU(2)_L$-doublet scalar $H_2$ and $SU(2)_L$-singlet charged scalar $\eta^\pm$, in addition to the SM-like Higgs doublet $H_1$.  Although the Wolfenstein version of the  model~\cite{Wolfenstein:1980sy} which assumes an additional  $Z_2$ symmetry, is ruled out by oscillation data~\cite{Koide:2001xy, He:2003ih}, it has been recently shown that the original version of the model~\cite{Zee:1980ai} is perfectly consistent with the neutrino oscillation data~\cite{Herrero-Garcia:2017xdu, Babu:2019mfe}. 

Following similar conventions, we adopt the scalar potential and the resulting  mass spectrum from Ref.\ \cite{Babu:2019mfe}. The cubic coupling ($\mu H_1^i H_2^j \eta^- \epsilon_{ij} + \rm h.c.$) in the scalar potential leads to mixing between $\eta^+$ and  $H_2^+$, with a mixing angle denoted as $\varphi$ and the physical charged Higgs states  denoted as $h^+$ and $H^+$, respectively.  Here, we work in a rotated basis for the Higgs doublets \cite{Babu:2018uik} in which only the neutral Higgs $H_1$ has a nonzero vacuum expectation value $v$. Now, one can express the Yukawa Lagrangian as 
\begin{eqnarray}
    \mathcal{L}_{y} & = & Y_{d}\bar{Q}_{L}d_{R}H_{1}+\widetilde{Y}_{d}\bar{Q}_{L}d_{R}H_{2}+Y_{u}\bar{Q}_{L}u_{R}\widetilde{H}_{1}\nonumber \\
     &  & +\widetilde{Y}_{u}\bar{Q}_{L}u_{R}\widetilde{H}_{2}+Y_{\ell}\bar{\psi}_{L}H_{1}\psi_{R}+\widetilde{Y}_{\ell}\bar{\psi}_{L}H_{2}\psi_{R} \\ \nonumber
     &  & +f\bar{\psi^c}_{L}\psi_{L}\eta^{+}+{\rm h.c.},\label{eq:model}
\end{eqnarray}
where the left-handed lepton and quark doublets are denoted as $\psi_{L}=(\nu, e)_{L}^{T}$ and $Q_{L}=(u, d)_{L}^{T}$.
\begin{figure}[t!]
    \centering
    \includegraphics[width=0.45 \textwidth]{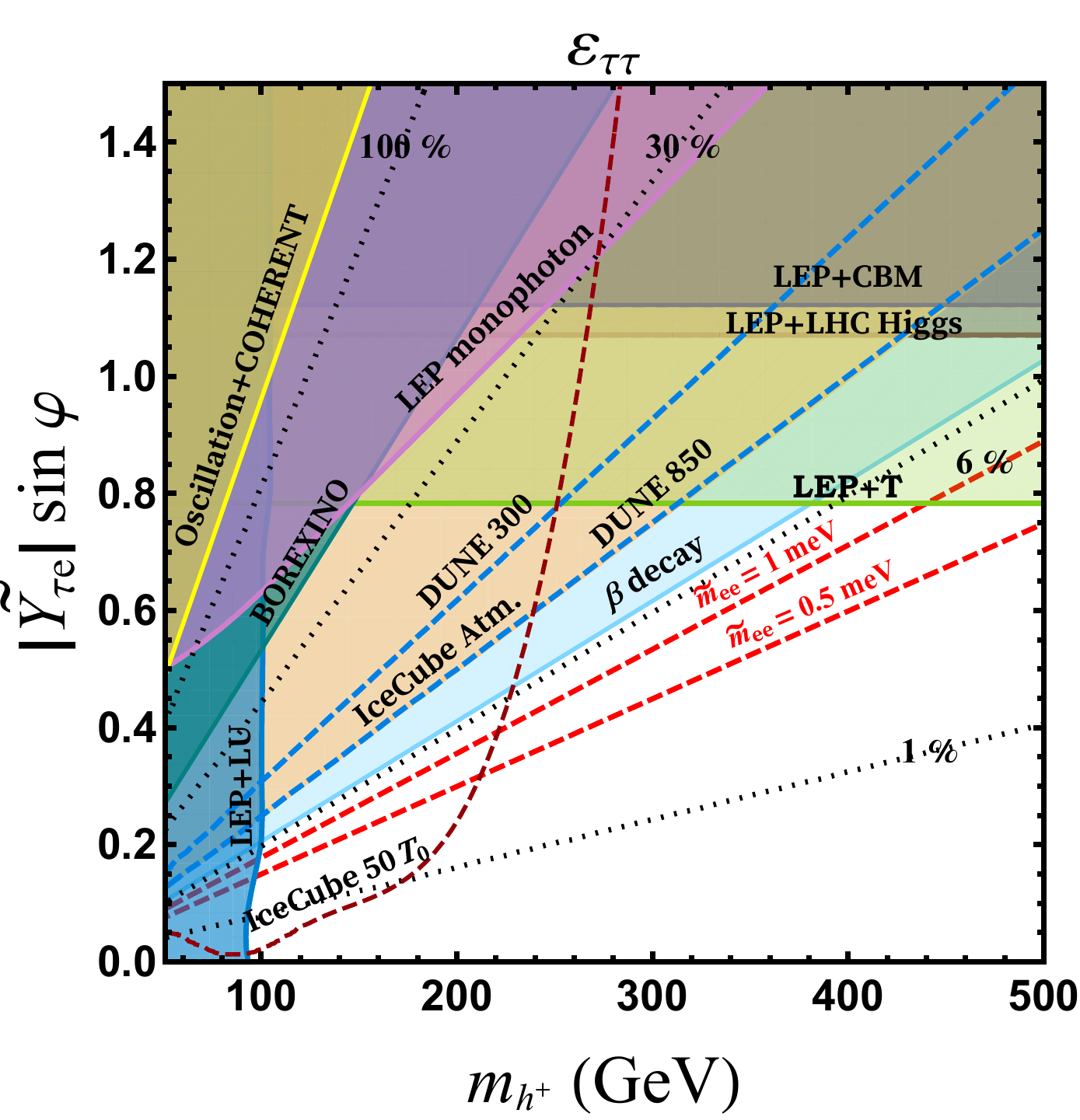}
    \caption{Implications of the charged Higgs in the Zee model. The red dashed lines are derived assuming non-observation of \ovbb in future experiments sensitive to $\tilde m_{ee}=1$ meV and 0.5 meV. Different color shaded regions are excluded from different experiments, black dotted lines depict the Zee model predictions for diagonal leptonic NSI $\varepsilon_{\tau \tau}$ and the other color dashed lines indicate the future sensitivity of different upcoming experiments; see text for details. 
    }
    \label{fig:NSI}
\end{figure}
Neutrino masses and mixings are generated radiatively at one-loop level as \cite{Zee:1980ai} $M_\nu  = \kappa  (f M_\ell Y_\ell + \widetilde{Y_\ell}^T \! M_\ell f^T )$, with the charged lepton mass matrix $M_\ell = Y_\ell v / \sqrt{2}$ and  $16 \pi^2 \kappa=  \sin{2 \varphi} \log\left(m_{h^+}^2 /m_{H^+}^2\right)$.
As we can see, the product of the  Yukawa couplings $f$ and $\widetilde{Y_\ell}$ is constrained from  neutrino oscillation data. 
In what follows, we will denote the entries of $Y_\ell$ and $\tilde Y_\ell$ as 
$Y_{\alpha \beta}$ and $\tilde Y_{\alpha \beta}$, respectively. 
In order to maximize the scalar interaction in the model~\cite{Babu:2019mfe}, we adopt the choice $\widetilde{Y_\ell}= {\cal O}(1)$ and $f\ll 1$ here. Since both Higgs doublets couple to up and down type quarks, a coupling  between the charged Higgs $h^+$ (which is mostly the singlet) and up-down type quarks will be induced via the  $\widetilde{Y}_q$ term  due to mixing with the charged Higgs $H^+$ (mostly the doublet).  Indeed, due to the presence of both  couplings $\widetilde{Y_\ell}$ and $\widetilde{Y}_q$, the scalar coupling  
$\epsilon_{\alpha}$ is generated by %{\bf WR: give indices of the Y!}
\begin{equation}
\epsilon_{\alpha} = \frac{1}{4 \sqrt{2} G_{F}}  \left(\frac{|\widetilde{Y}_{\alpha e} \widetilde{Y}_{ud}|\sin ^{2} \varphi}{m_{h^{+}}^{2}}+\frac{|\widetilde{Y}_{\alpha e}\widetilde{Y}_{ud}|\cos ^{2} \varphi}{m_{H^{+}}^{2}}\right).
\end{equation}
Moreover, the leptonic Yukawa $\widetilde{Y_\ell}$ itself leads to  an  effective four-fermion Lagrangian relevant for vector NSI  in the form  $$\mathcal{L}_{\mathrm{eff}}=\left(\bar{\nu}_{ \tau} e_{R}\right)\left(\bar{e}_{R} \nu_{ \tau}\right)\left(\frac{|\widetilde{Y}_{\tau e} |^2\sin ^{2} \varphi}{m_{h^{+}}^{2}}+\frac{|\widetilde{Y}_{\tau e} |^2\cos ^{2} \varphi}{m_{H^{+}}^{2}}\right),$$ 
where we consider the $\tau$ flavour of neutrino ($\widetilde{Y}_{\tau e}\neq 0$) for simplicity and to avoid other stringent limits from  lepton universality tests in $W$ decays~\cite{LEP:2003aa} and Michel parameter constraints \cite{Tanabashi:2018oca}. Applying the Fierz transformation, we can rewrite it as $$\mathcal{L}_{\mathrm{eff}}=-\frac{1}{2}\left(\bar{\nu}_{ \tau} \gamma_{\mu} \nu_{ \tau}\right)\left(\bar{e}_{R} \gamma^{\mu} e_{R}\right)\left(\frac{|\widetilde{Y}_{\tau e} |^2\sin ^{2} \varphi}{m_{h^{+}}^{2}}+\frac{|\widetilde{Y}_{\tau e} |^2\cos ^{2} \varphi}{m_{H^{+}}^{2}}\right),$$ 
which can be directly compared to the standard parametrization of NSI in the form \cite{Wolfenstein:1977ue} 
$$\mathcal{L}_{\mathrm{NSI}}^{\mathrm{NC}}=-2 \sqrt{2} G_{F} \sum_{f, X, \alpha, \beta} \varepsilon_{\alpha \beta}^{f X}\left(\bar{\nu}_{\alpha} \gamma^{\mu} P_{L} \nu_{\beta}\right)\left(\bar{f} \gamma_{\mu} P_{X} f\right).$$ 
Hence, at the tree level,  the leptonic NSI induced by the singly-charged scalars $h^{\pm},H^{\pm} $  is given by
\begin{equation}
\varepsilon_{\tau \tau}^{eR}
\equiv \varepsilon_{\tau \tau} 
=\frac{1}{4 \sqrt{2} G_{F}} \left(\frac{|\widetilde{Y}_{\tau e} |^2\sin ^{2} \varphi}{m_{h^{+}}^{2}}+\frac{|\widetilde{Y}_{\tau e} |^2\cos ^{2} \varphi}{m_{H^{+}}^{2}}\right).
\end{equation}
 In the quark sector, even if one starts with a diagonal texture of Yukawa matrices $\widetilde{Y}_q$, off-diagonal entries reappear due to non-vanishing CKM entries.  This is why we assume all the scalars originating primarily from the second doublet reside above the TeV scale and that all entries except the {(1,1)} one in both the up-type and down-type quark Yukawa matrices are sufficiently small in order to avoid flavor-changing neutral current (FCNC) processes. However, there will be still strong constraints from  meson decays, CKM unitarity, the $T$ parameter, beta decay, etc. As  mentioned earlier, the most stringent bound  originates from beta decay combined with meson decay data, which reads $\epsilon_{\tau} <0.063$.  In the lepton sector, there will be stringent constraints from charged lepton flavour violating processes. However, we make such judiciary choice in Yukawa matrices $\widetilde{Y_l}$ ($\widetilde{Y}_{\tau e}\neq 0, Y_{\alpha \tau} \neq 0$ for $\alpha=e$ or $\mu$) that no more than one entry in a given row of $\widetilde{Y_\ell}$ is large. Such a choice (keeping only the entries $(\widetilde{Y}_{ee}, \, \widetilde{Y}_{\mu e}, \, \widetilde{Y}_{\tau e})$ one at a time in the first column of $\widetilde{Y_\ell}$)  does not lead to $\ell_\alpha \rightarrow \ell_\beta + \gamma$ decay mediated by the charged scalar and also does not affect the maximum strength of the NSI. These choices also automatically satisfy LEP contact interaction limits for the process $e^- e^+ \to f \bar{f}$ \cite{Babu:2019mfe}.  With this choice, the charged scalar, $h^+$, can be as light as $\sim 110$ GeV, while satisfying all these experimental constraints \cite{Babu:2019mfe}. The LHC contact interaction limits are also not directly applicable if the charged scalar $h^+$ is that light. Further note that, our choice of parameter space leads to the dominant decay mode of $h^+ \to \tau \nu$, and thus, it can escape the stringent LHC constraint imposed by analyses looking for charged lepton ($e$ or $\mu$) and missing transverse momentum events. In a Nevertheless, there will be still several  theoretical and experimental constraints, such as charge breaking minima, electroweak precision tests, charged lepton flavour violation, collider constraints from LEP and LHC, lepton universality tests and monophoton constraints,  all of which are shown in Fig.~\ref{fig:NSI}. The yellow, dark-green and orange shaded regions depict the direct constraints on  $\varepsilon_{\tau \tau}$ from  a global fit to neutrino oscillation + COHERENT data~\cite{Esteban:2018ppq}, neutrino-electron scattering experiments such as Borexino~\cite{Agarwalla:2019smc}, and IceCube atmospheric neutrino data~\cite{Esmaili:2013fva}, respectively. The purple shaded region is excluded by the monophoton process $e^+ e^- \to \nu\bar{\nu}\gamma$ from LEP data~\cite{Berezhiani:2001rs}, since the  $ee\nu\nu$ operator leads to an additional contribution for the process mediated by $h^+$ in the  $t$-channel. In order to maximize leptonic NSI,  a large mixing between the singlet and doublet charged scalar fields, and hence, a large trilinear $\mu$ term needs to be introduced.  However, it cannot be arbitrarily large, since it leads to charge breaking minima of the potential. This limit is shown by the gray shaded region. A light charged scalar additionally contributes  to the $h \to \gamma \gamma $ process and the limit from LHC Higgs data \cite{ATLAS:2020qdt} is shown as the brown shaded region. The light green colored region is excluded by the $T$ parameter \cite{Tanabashi:2018oca}. LEP constraints on charged scalar searches are shown by the blue shaded region. The blue dashed lines indicate the future DUNE sensitivities on $\epsilon_{\tau\tau}$ for 300 and 850 kt MW yr exposure~\cite{dev_pondd}.  At  IceCube, the light charged scalar $h^+$ of the model could potentially lead to a Glashow-like resonance feature \cite{Babu:2019vff} in the ultra-high energy neutrino event spectrum; the dashed brown curve indicates this future IceCube sensitivity  corresponding to an exposure time of $50 \times T_{0}$ \cite{Babu:2019vff}, where $T_0=$ 2653 days. For completeness of our study,  we set $\widetilde{Y}_{\tau e}=\widetilde{Y}_{u d} $ in Fig.~\ref{fig:NSI} to  find a direct correlation between $\epsilon_\tau$ and $\varepsilon_{\tau \tau}$. The cyan region is excluded from the $\beta$ decay limit.  The red dashed lines are derived assuming non-observation of \ovbb decay in future experiments sensitive to  $\tilde m_{ee}$ of 1 meV and 0.5~meV.

\section{\label{sec:con}Conclusion}
\noindent
We have pointed out that neutrinoless double beta decay can be mediated by a general $m_{\alpha \beta}$ mass-insertion, instead of the usually considered standard mechanism with its $m_{ee}$ mass-insertion. This can be realized via new lepton number conserving scalar interactions.

The funnel of neutrinoless double beta decay of infinite half-life can be sealed in this way, which we demonstrated explicitly. A UV completion of the new scalar interaction within the Zee model of radiative neutrino mass generation was shown to be in agreement with all existing limits from neutrino oscillations to collider physics. 

While the absence of the funnel can be expected in any scenario with more than one contribution to \ovbb decay, the new mechanism discussed here depends on the lightest neutrino mass. Lepton number violation is only implied by the Majorana mass term of the light neutrinos.

\appendix
\section{Derivation of the \ovbb Decay Half-Life}
\label{app:a}
\subsection{Particle Physics}
\noindent
    In the presence of the new interactions in Eq.~(\ref{eq:m}), the
    amplitude of $0\nu\beta\beta$ decay is modified to 
    \begin{equation}
    i{\cal M}_{V+S}^{(0\nu\beta\beta)}=\langle e_{1}e_{2}F|({\cal L}_{V}+{\cal L}_{S})({\cal L}_{V}+{\cal L}_{S})|I\rangle,\label{eq:m-17}
    \end{equation}
    where $\langle F|$ and $|I\rangle$ are the final and initial nuclear
    states, and $\langle e_{1}e_{2}|$ denotes the final electron states.
    Expanding Eq.~(\ref{eq:m-17}), we obtain
    \begin{widetext}
    \begin{equation}
    i{\cal M}_{V+S}^{(0\nu\beta\beta)}=(8G_{F}^{2}V_{ud}^{2})\int\frac{d^{4}q}{(2\pi)^{4}}\left\{ \left[i{\cal M}_{S}^{({\rm lep})}\right]\left[i{\cal M}_{S}^{({\rm nuc})}\right]+\left[i{\cal M}_{SV}^{({\rm lep})}\right]^{\mu}\left[i{\cal M}_{SV}^{({\rm nuc})}\right]_{\mu}+\left[i{\cal M}_{V}^{({\rm lep})}\right]^{\mu\nu}\left[i{\cal M}_{V}^{({\rm nuc})}\right]_{\mu\nu}\right\} ,\label{eq:m-18}
    \end{equation}
    where ${\cal M}^{({\rm lep})}_X$ and ${\cal M}^{({\rm nuc})}_X$ are
    the leptonic and nuclear parts of the amplitude, and $q$ is the internal
    momentum. The subscripts $S$, $V$, and $SV$ of the partial amplitudes
    indicate that they are from ${\cal L}_{S}{\cal L}_{S}$, ${\cal L}_{V}{\cal L}_{V}$,
    and ${\cal L}_{V}{\cal L}_{S}$ or ${\cal L}_{S}{\cal L}_{V}$ in
    Eq.~(\ref{eq:m-17}).

    The leptonic amplitudes read
    \begin{eqnarray}
    i{\cal M}_{S}^{({\rm lep})} & = & \overline{u_{e}}P_{L}\frac{i}{\slashed{q}}(-i\epsilon_{\alpha}\thinspace m_{\alpha\beta}\thinspace\epsilon_{\beta})\frac{i}{\slashed{q}}P_{L}u_{e^{c}} 
    %\nonumber \\ & = & 
    = \overline{u_{e}}\frac{im_{\alpha\beta}\thinspace\epsilon_{\alpha}\thinspace\epsilon_{\beta}}{q^{2}}P_{L}u_{e^{c}},\label{eq:m-19}
    \end{eqnarray}
    \begin{eqnarray}
    i{\cal M}_{SV}^{({\rm lep})} & = & \overline{u_{e}}\gamma^{\mu}P_{L}\frac{i}{\slashed{q}}(-i\thinspace m_{e\beta}\thinspace\epsilon_{\beta})\frac{i}{\slashed{q}}P_{L}u_{e^{c}} 
    %\nonumber \\ &  & 
     +\overline{u_{e}}P_{L}\frac{i}{\slashed{q}}(-i\epsilon_{\alpha}\thinspace m_{\alpha e})\frac{i}{\slashed{q}}P_{L}\gamma^{\mu}u_{e^{c}} 
     %\nonumber \\     & = &  
     = \overline{u_{e}}\frac{im_{e\beta}\thinspace\epsilon_{\beta}}{q^{2}}\gamma^{\mu}u_{e^{c}},\label{eq:m-20}
    \end{eqnarray}
    \begin{eqnarray}
    i{\cal M}_{V}^{({\rm lep})} & = & \overline{u_{e}}\gamma^{\mu}P_{L}\frac{i}{\slashed{q}}(-im_{ee})\frac{i}{\slashed{q}}P_{L}\gamma^{\nu}u_{e^{c}} 
    %\nonumber \\     & = &   
    = \overline{u_{e}}\frac{im_{ee}}{q^{2}}\gamma^{\mu}\gamma^{\nu}P_{R}u_{e^{c}},\label{eq:m-21}
    \end{eqnarray}
    where $q$ is the neutrino momentum, $\overline{u_{e}}$ and $u_{e^{c}}$
    are the two electron final states, and $m_{\alpha\beta}$ is an element of the neutrino
    mass matrix. After antisymmetrizing the above three expressions in the electron wave functions (i.e.\ taking both $t$ and $u$ diagrams into account), the second one, combining the SM vector and BSM scalar vertices, vanishes. The corresponding contribution could be in principle still saved by a compensating minus sign coming from the nuclear part of the amplitude. However, this would require inclusion of electron $P$ wave, and thus would result in a subleading contribution, which we neglect. We explicitly checked this does not affect our conclusions.
    
    The relevant nuclear amplitudes, defined as
    \begin{eqnarray}
    i{\cal M}_{S}^{({\rm nuc})} & \equiv & \langle F|\overline{u}P_{L}d\overline{u}P_{L}d|I\rangle,\label{eq:m-22}\\
    i{\cal M}_{V}^{({\rm nuc})} & \equiv & \langle F|\overline{u}\gamma^{\mu}P_{L}d\overline{u}\gamma^{\nu}P_{L}d|I\rangle,\label{eq:m-24}
    \end{eqnarray}
    give rise, after applying the hadronization procedure (described e.g.\ in Ref.~\cite{Graf:2018ozy}), to the nuclear matrix elements $M_{S}$ and $M_{V}$ corresponding to the purely scalar and the standard vector contribution, respectively. The structure of these is discussed in more detail in the next subsection.

    Combining the leptonic and nuclear amplitudes, we obtain 
    \begin{equation}
    i{\cal M}_{V+S}^{(0\nu\beta\beta)}=i\overline{u_{e}}\left[\lambda_{V}P_{R}+\lambda_{S}P_{L}\right]u_{e^{c}}\thinspace,\label{eq:m-28}
    \end{equation}
    where
    \begin{equation}
    (\lambda_{V},\ \lambda_{S})\propto(m_{ee}M_{V},\ m_{\alpha\beta}\thinspace\epsilon_{\alpha}\thinspace\epsilon_{\beta}M_{S}).\label{eq:m-29}
    \end{equation}
    Applying the trace technology, we obtain the following squared amplitude \begin{eqnarray}
     %&  & 
     \left|i{\cal M}_{V+S}^{(0\nu\beta\beta)}\right|^{2}
     % \nonumber \\ 
     & = & 
     {\rm Tr}\left\{ u_{e}\overline{u_{e}}\left[\lambda_{V}P_{R}+\lambda_{S}P_{L}\right] 
     %\right.\nonumber \\&  & \times\left.  
     u_{e^{c}}\overline{u_{e^{c}}}\left[\lambda_{V}^{*}P_{L}+\lambda_{S}^{*}P_{R}\right]\right\} \nonumber \\
     & = & 2p_{1}\cdot p_{2}\left[|\lambda_{V}|^{2}+|\lambda_{S}|^{2}\right]
     +2m_{e}^{2}\left[\lambda_{S}\lambda_{V}^{*}+\lambda_{V}\lambda_{S}^{*}\right].\label{eq:m-4}
    \end{eqnarray}
    Here, we are considering $S_{1/2}$ approximation of the electron wave functions.
    We have used $u_{e}\overline{u_{e}}=\slashed{p}_{1}+m_{e}$
    and $u_{e^{c}}\overline{u_{e^{c}}}=\slashed{p}_{2}+m_{e}$, with $p_{1}$
    and $p_{2}$ being the two momenta of the outgoing electrons.

    The decay rate of $0\nu\beta\beta$ is computed by (see the appendix
    of Ref.~\cite{Deppisch:2020sqh})
    \begin{eqnarray}
    \Gamma_{0\nu} & = & \int\frac{\left|i{\cal M}_{V+S}^{(0\nu\beta\beta)}\right|^{2}2\pi\delta}{4M_{N}(M_{N}-\Delta M)}\left[\prod_{i=1}^{2}\frac{d^{3}p_{i}}{(2\pi)^{3}2E_{i}}\right],\label{eq:m-31}
    \end{eqnarray}
    where $M_{N}$ is the mass of the initial nucleus, $\Delta M$ is
    the mass difference between the final and the initial nuclei, and
    \begin{equation}
    \delta=\delta(\Delta M-E_{1}-E_{2}).\label{eq:m-10}
    \end{equation}
    Note that in the above phase space integral, the spatial part of $p_{1}\cdot p_{2}$
    in $\left|i{\cal M}_{V+S}^{(0\nu\beta\beta)}\right|^{2}$ does not
    contribute because
    \begin{eqnarray}
      \int p_{1}\cdot p_{2}\prod_{i=1}^{2}\frac{d^{3}p_{i}}{(2\pi)^{3}2E_{i}}\delta %\nonumber \\ & = & 
     \int\left(E_{1}E_{2}-|\vecs{p_{1}}||\vecs{p_{2}}|\cos\theta\right)\frac{|\vecs{p_{1}}|^{2}|\vecs{p_{2}}|^{2} d |\vecs{p_{1}}| d|\vecs{p_{2}}|d\cos\theta}{(2\pi)^{4}2E_{1}E_{2}}\delta
     %\nonumber \\     & = & 
     = \int E_{1}E_{2}\frac{|\vecs{p_{1}}|^{2}|\vecs{p_{2}}|^{2} d |\vecs{p_{1}}| d|\vecs{p_{2}}|}{(2\pi)^{4}E_{1}E_{2}}\delta,\nonumber
    \end{eqnarray}
    where $\theta$ is the angle between $\vecs{p}_{1}$ and $\vecs{p}_{2}$.
    The spatial part vanishes when $\theta$ is integrated out.

    Now we define
    \begin{equation}
    R\equiv\frac{\int_{p}m_{e}^{2}}{\int_{p}E_{1}E_{2}}, \label{eq:m-30}
    \end{equation}
    with the notation  $\int_{p}F\equiv\int F\frac{|\vecs{p_{1}}|^{2}|\vecs{p_{2}}|^{2} d |\vecs{p_{1}}| d|\vecs{p_{2}}|}{(2\pi)^{4}E_{1}E_{2}}\delta$
    for an arbitrary function $F$.  Then, in order to compute Eq.~(\ref{eq:m-31}),
    one simply needs to replace $p_{1}\cdot p_{2}\rightarrow\int_{p}E_{1}E_{2}$,
    $E_{1}E_{2}\rightarrow\int_{p}E_{1}E_{2}$, $m_{e}^{2}\rightarrow\int_{p}m_{e}^{2}$
    in Eq.~(\ref{eq:m-4}).

    Therefore, our final result of $0\nu\beta\beta$
    decay rate including the new interaction reads
    \begin{align}
    \Gamma_{0\nu} &= \frac{G_{0\nu}}{m_{e}^{2}}\left[\left|m_{ee}M_{V}\right|^{2}+\left|m_{\alpha\beta}\thinspace\epsilon_{\alpha}\epsilon_{\beta}M_{S}\right|^{2}\right] +\frac{G_{0\nu}}{m_{e}^{2}}2R\thinspace{\rm Re}\left[m_{\alpha\beta}\thinspace\epsilon_{\alpha}\epsilon_{\beta}M_{S}m_{ee}^{*}M_{V}^{*}\right].\label{eq:m-13}
    \end{align}
    Here $R$, defined in Eq.~(\ref{eq:m-30}), is referred
    to as {\it interference factor}.  Given an isotope
    dependent $Q$, with  $\Delta M=Q+2m_{e}$, $R$ can be
    evaluated numerically according to Eq.~(\ref{eq:m-30}). In Fig.~\ref{fig:Q},
    we present the $R(Q)$ curve and indicate the
    values for several typical isotopes. 

\subsection{Nuclear Matrix Elements}
\label{app:b}
\noindent 
For the derivation of the relevant nuclear matrix elements (NMEs) entering the rate of the studied \ovbb decay contributions we follow the standard procedure described in detail e.g.\ in Ref.~\cite{Graf:2018ozy}. Hence, after parametrizing the nucleon matrix elements of the considered quark bilinears in terms of the nucleon form factors  $F_X(q)$ we make use of the nonrelativistic expansion and obtain an approximate expression for individual currents, combinations of which give the actual NMEs. As usual, we consider only $0^+ \rightarrow 0^+$ transition.

For the standard mechanism we use the usual expression (see e.g.~\cite{Deppisch:2020ztt}, where the same notation was used) giving the total NME $M_V$ as
\begin{align}
\label{eq:nme_nu}
	M_V &= g_V^2 \nme_F - g_A^2 \nme_{GT}^{AA} 
	       + \frac{g_A g_{P}}{6} 
	         \left(\nme^{\prime AP}_{GT} + \nme^{\prime AP}_T\right) \\
          &+ \frac{(g_V+g_W)^2}{12}
             \left(-2\nme^{\prime WW}_{GT} + \nme^{\prime WW}_T\right) 
             %\\           & 
             - \frac{g_{P}^2}{48} 
             \left(\nme^{\prime\prime PP}_{GT} 
                 + \nme^{\prime\prime PP}_T
             \right).\nonumber
\end{align}
Here, the form factor charges $g_X$ correspond to the value of the nucleon form factors $F_X(q)$ at zero momentum transfer, i.e.\ $g_X = F_X(0)$. We employ the following numerical values: $g_V=1$, $g_A=1$, $g_W=3.7$, $g_P=231$~\cite{Simkovic:1999re}, $g_S=1$~\cite{Gonzalez-Alonso:2018omy}, $g_{P'}=349$~\cite{Gonzalez-Alonso:2018omy}. We assume a quenching of the effective nuclear axial coupling, as often employed in the literature, 
see e.g.\ the discussion in Ref.~\cite{Deppisch:2020ztt}. The large values of the pseudoscalar couplings $g_P$ and $g_{P'}$ are due to enhancement by the implicit pion resonance.
The dependence on the momentum $q$ transferred between the two beta-decaying nucleons captured by the product of the reduced form factors $F_X(q^2)/g_X$ is then incorporated within the elementary nuclear matrix elements entering Eq.~\eqref{eq:nme_nu}. Table~\ref{tab:nmes} shows the explicit form of the individual Fermi ($M_F$), Gamow-Teller ($M_{GT}$) and tensor ($M_T$) NMEs including the corresponding reduced form factor products $\tilde h(q^2)$.

\renewcommand{\arraystretch}{2}
\setlength{\tabcolsep}{12pt}
\begin{table*}[t!]
\centering
\begin{tabular}{ll}
\toprule
\text{\emph{Standard Mechanism}} \\
\botrule
NME & $\tilde{h}_\circ(q^2)$ \\
\cline{1-1} \cline{2-2}
$M_F = \langle h_{XX}(q^2) \rangle$ & $\tilde{h}_{XX}(q^2) = \frac{1}{(1+q^2/m_V^2)^4}$ \\
$M^{\prime WW}_{GT} = 
\left\langle \frac{\vecl{q}^2}{m_p^2} h_{XX}(q^2) 
(\vecs{\sigma}_a\cdot\vecs{\sigma}_b) \right\rangle$		
& $\tilde{h}_{XX}(q^2)$                                         \\
$M^{\prime WW}_T = 
\left\langle \frac{\vecl{q}^2}{m_p^2} h_{XX}(q^2) \vecl{S}_{ab} \right\rangle$	
& $\tilde{h}_{XX}(q^2)$                                         \\
$M_{GT}^{AA} = \langle h_{AA}(q^2) (\vecs{\sigma}_a\cdot\vecs{\sigma}_b) \rangle$	
& $\tilde{h}_{AA}(q^2) = \frac{1}{\left(1+q^2/m_A^2\right)^4}$  \\
$M^{\prime AP}_{GT} =  
\left\langle \frac{\vecl{q}^2}{m_p^2} h_{AP}(q^2) 
(\vecs{\sigma}_a\cdot\vecs{\sigma}_b) \right\rangle$ 
& $\tilde{h}_{AP}(q^2) = \frac{1}{\left(1 + q^2/m_A^2\right)^4}
\frac{1}{1 + q^2/m_\pi^2}$                                      \\
$M^{\prime AP}_T = 
\left\langle \frac{\vecl{q}^2}{m_p^2}  h_{AP}(q^2) \vecl{S}_{ab} \right\rangle$	
& $\tilde{h}_{AP}(q^2)$                                         \\
$M^{\prime\prime PP}_{GT} = \left\langle \frac{\vecl{q}^4}{m_p^4}
h_{PP}(q^2) (\vecs{\sigma}_a\cdot\vecs{\sigma}_b) \right\rangle$	
& $\tilde{h}_{PP}(q^2)=\frac{1}{(1 + q^2/m_A^2)^4}
\frac{1}{(1 + q^2/m_\pi^2)^2}$                                         \\
$M^{\prime\prime PP}_{T} = \left\langle \frac{\vecl{q}^4}{m_p^4}  
h_{PP}(q^2) \vecl{S}_{ab} \right\rangle$ 
& $\tilde{h}_{PP}(q^2)$                                         \\
\botrule
\text{\emph{Flavoured Scalar Mechanisms}} \\
\botrule
NME & $\tilde{h}_\circ(q^2)$ \\
\cline{1-1} \cline{2-2}
$M_{GT}^{'P'P'} = \left\langle \frac{\vecl{q}^2}{m_p^2} h_{PP}(q^2) (\vecs{\sigma}_a\cdot\vecs{\sigma}_b) \right\rangle$ & $\tilde{h}_{PP}(q^2)$ \\
$M_{T}^{'P'P'} = \left\langle \frac{\vecl{q}^2}{m_p^2} h_{PP}(q^2) \vecl{S}_{ab} \right\rangle$ & $\tilde{h}_{PP}(q^2)$ \\
\botrule
\end{tabular}
\caption{\label{tab:nmes}Here we list the double beta decay Fermi ($M_F$), Gamow-Teller ($M_{GT}$) and tensor ($M_T$) NMEs entering Eqs.~\eqref{eq:nme_nu} and~\eqref{eq:nmeS}. The relevant reduced form factor product $\tilde{h}(q^2)$ is always shown alongside. These $q$-dependent functions enter the NMEs multiplied by the neutrino potential given in Eq.~\eqref{eq:neutrinopotentialLR}; hence we define, $h_\circ(q^2) = v(q^2)\tilde h_\circ(q^2)$. The label $X$ in the above definitions stands collectively for vector, weak-magnetism and tensor form factors involved, i.e.\ $X = V, W, T$, as all of these come with the same shape parameter $m_V=0.84$ GeV~\cite{Schindler:2006jq}. The $q$ dependencies of the axial and pseudoscalar form factors denoted by subscripts $A$ and $P$, respectively, include the shape parameter $m_A=1.09$ GeV~\cite{Schindler:2006jq} and for the pion mass we take $m_{\pi}=0.138$~GeV. The spin operators corresponding to individual nucleons $a$, $b$ are represented using the Pauli matrices $\vecs{\sigma}_{a,b}$ and the tensor NMEs incorporate the operator $S_{ab} = 3(\vecs{\sigma}_a \cdot \vecl{q}) (\vecs{\sigma}_b \cdot \vecl{q}) - (\vecs{\sigma}_a \cdot \vecs{\sigma}_b)$. 
}
\end{table*}
All the NMEs defined in Tab.~\ref{tab:nmes} contain on top of the product of the reduced nucleon form factors $\tilde{h}(q^2)$ also the neutrino potential, which captures the $q$ dependence of the long-range exchange of an essentially massless neutrino mediating the $0\nu\beta\beta$ decay. As we follow the formulation of \cite{Simkovic:1999re} and \cite{Barea:2013bz}, the two-body transition operator is constructed in momentum space. The neutrino potential for the studied mechanisms reads
\begin{align}
\label{eq:neutrinopotentialLR}
	v(q) = \frac{2}{\pi}\frac{1}{q(q + \tilde A)}.
\end{align}
Here, given the typical neutrino momentum $q$, the neutrino mass has been neglected and $\tilde{A}$ denotes the closure energy, taken from Ref.\  \cite{Haxton:1985am} or estimated by the systematics, and finally $\tilde{A}=1.12A^{1/2}$ MeV.

In the same spirit we derive the NMEs contributing to the scalar \ovbb decay mechanism discussed here. The product of two scalar nucleon currents can be expressed as
\begin{align}
    \label{eq:nmeS}
    \nme_S = g_S^2\nme_F + \frac{g_{P'}^2}{12}\left(\nme_{GT}^{'P'P'}-\nme_T^{'P'P'}\right),
\end{align}
where the elementary NMEs on the right-hand side are defined in Tab.~\ref{tab:nmes}.

To calculate the numerical values of the derived NMEs a nuclear structure model must be employed. Here, we make use of so called Interacting Boson Model (IBM-2) \cite{Barea:2009zza, Barea:2013bz, Barea:2015kwa}.
 
\section{Magnitudes of neutrino mass matrix elements}
\renewcommand{\thefigure}{B\arabic{figure}}
\setcounter{figure}{0}

\noindent
In addition to $m_{ee}$, neutrino mass matrix elements of other flavours are also involved in this work. 
It is therefore useful to know their magnitudes and dependence on the smallest mass $m_{L}$, which have been studied in the literature -- see  \cite{Merle:2006du,Grimus:2012ii}.
For the reader's convenience (and also as an update), in this appendix we display in Fig.~\ref{fig:mab} the magnitudes of all neutrino mass matrix elements.

    \begin{figure}[h]
        \centering
        \includegraphics[width=0.8 \textwidth]{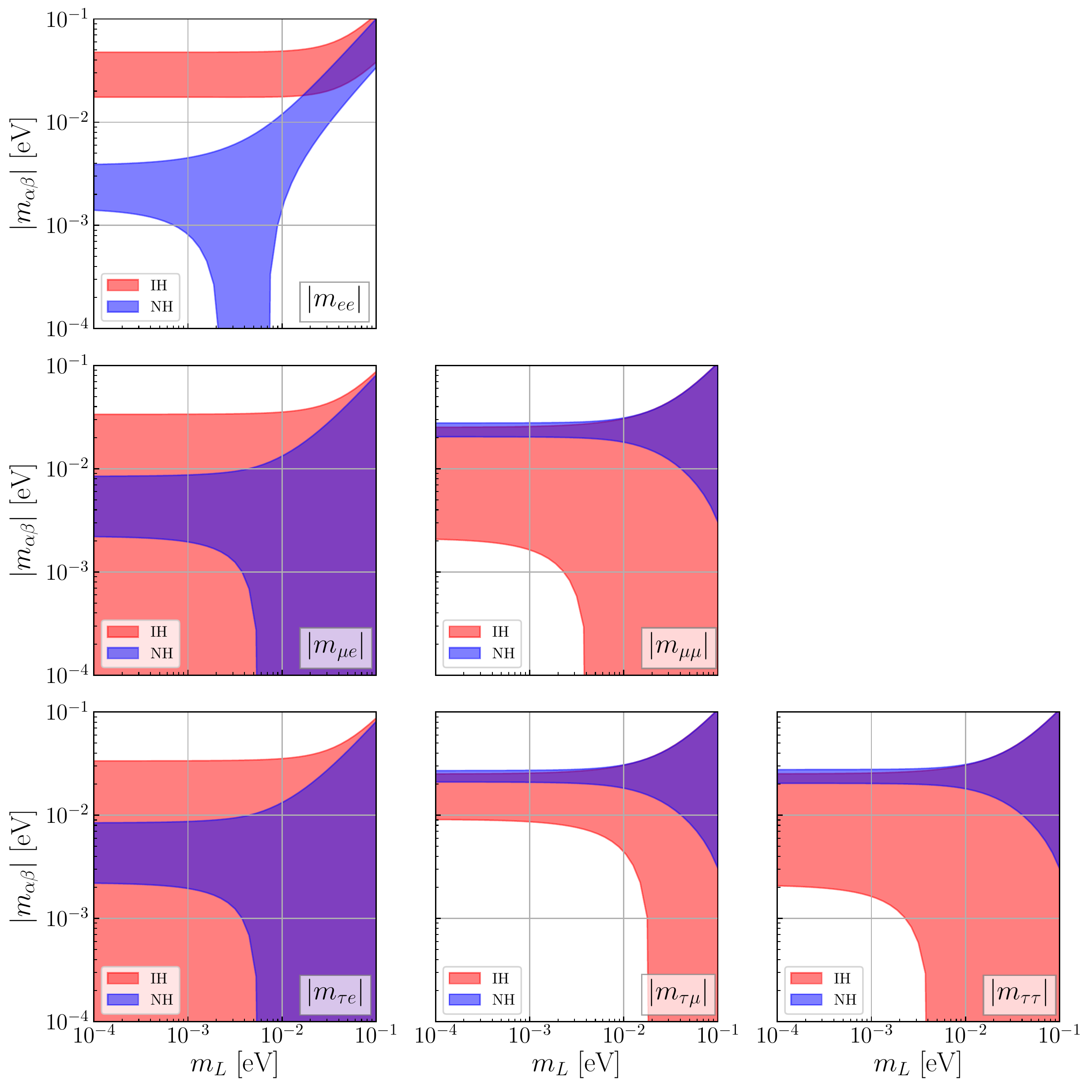}
        \caption{The individual neutrino mass matrix elements against the smallest neutrino mass $m_{L}$, see also \cite{Merle:2006du,Grimus:2012ii}. Note the funnel appearing in the $m_{ee}$ case.}
        \label{fig:mab}
    \end{figure}

\begin{acknowledgments}
The authors would like to thank Frank F.\ Deppisch and Oliver Scholer for useful comments. LG is grateful to Jose Barea for providing the code to calculate the standard
mechanism of double beta decay in the IBM-2 framework.
\end{acknowledgments}

\end{widetext}
\clearpage
\bibliographystyle{utphys}
\bibliography{reference}

\end{document}